\documentclass[useAMS,usenatbib,usegraphicx]{mn2e}
  
\usepackage{aas_macros}

\title[Disc formation in turbulent massive cores]{Disc formation in turbulent massive cores: Circumventing the magnetic braking catastrophe}
  \author[D. Seifried et al.]
  {D.~Seifried,$^{1,2,3}$\thanks{dseifried@hs.uni-hamburg.de} R.~Banerjee,$^{1}$ R.~E.~Pudritz,$^3$ R.~S.~Klessen$^2$ \\
  $^1$Hamburger Sternwarte, Universit\"at Hamburg, Gojenbergsweg 112, 21029 Hamburg, Germany\\
  $^2$Institut f\"ur Theoretische Astrophysik, Universit\"at Heidelberg, Albert-Ueberle-Str. 2, 69120 Heidelberg, Germany \\
  $^3$Department of Physics $\&$ Astronomy, McMaster University, Hamilton ON L8S 4M1, Canada}

\date{Accepted 2012 March 7.  Received 2012 March 7; in original form 2012 January 26}

\pagerange{\pageref{firstpage}--\pageref{lastpage}} \pubyear{2012}

\begin{document}

\label{firstpage}

\maketitle

\begin{abstract}
We present collapse simulations of 100 M$_{\sun}$, turbulent cloud cores threaded by a strong magnetic field. During the initial collapse phase filaments are generated which fragment quickly and form several protostars. Around these protostars Keplerian discs with typical sizes of up to 100 AU build up in contrast to previous simulations neglecting turbulence. We examine three mechanisms potentially responsible for lowering the magnetic braking efficiency and therefore allowing for the formation of Keplerian discs. Analysing the condensations in which the discs form, we show that the build-up of Keplerian discs is neither caused by magnetic flux loss due to turbulent reconnection nor by the misalignment of the magnetic field and the angular momentum. It is rather a consequence of the turbulent surroundings of the disc which exhibit no coherent rotation structure while strong local shear flows carry large amounts of angular momentum. We suggest that the "magnetic braking catastrophe", i.e. the formation of sub-Keplerian discs only, is an artefact of the idealised non-turbulent initial conditions and that turbulence provides a natural mechanism to circumvent this problem.
\end{abstract}

\begin{keywords}
 MHD -- methods: numerical -- stars: formation -- accretion discs
\end{keywords}

\section{Introduction}

In recent years a great number of simulations have been performed that investigate the formation of protostellar discs under the influence of magnetic fields~\citep[e.g.][]{Allen03,Matsumoto04,Machida05,Banerjee06,Banerjee07,Price07,Hennebelle08,Hennebelle09,Duffin09,Commercon10,Peters11,Seifried11a}. In simulations with magnetic field strengths comparable to observations~\citep[e.g.][]{Falgarone08,Girart09,Beuther10} no rotationally supported discs were found. As strong magnetic braking is responsible for the removal of the angular momentum, this problem is also called the ''magnetic braking catastrophe''.
The results of these numerical simulations stand in sharp contrast to observations, which show that discs are present in the earliest stage of protostellar evolution~\citep[e.g.][]{Jorgensen09}.
It is also well known that outflows are the first observable signatures of star formation, and that therefore discs should already have formed~\citep[e.g.][]{Arce07}.

The inclusion of ambipolar diffusion~\citep{Mellon09,Duffin09} also fails to produce Keplerian discs in the earliest evolutionary stages. Considering the effect of Ohmic dissipation, only very small ($\sim$ 10 solar radii) Keplerian discs were found~~\citep{Dapp11a,Dapp11b}, unless an unusually high resistivity is used~\citep{Krasnopolsky10}. However, recently two mechanisms solving the catastrophic magnetic braking problem were proposed, i.e. the inclusion of the Hall effect~\citep{Krasnopolsky11} and turbulent reconnection~\citep{Santos12}.

In this paper we remedy a shortcoming of the simulations referred to in the beginning namely their lack of turbulent motions. We present results from a number of simulations investigating the possible role of turbulence in reducing the magnetic braking efficiency and allowing for the formation of protostellar discs.

\section{Initial conditions} \label{sec:techniques}

We now shortly describe the basic simulation setup. For a more detailed description we refer the reader to~\citet{Seifried11a}. We simulate the collapse of a 100 M$_{\sun}$ molecular cloud core, 0.25 pc in size and embedded in 0.75 pc sized cubic simulation box of low-density gas ($4.2 \times 10^{-21}$  g cm$^{-3}$). The density in the core declines as $\rho \propto r^{-1.5}$ having a maximum of $2.3 \times 10^{-17}$ g cm$^{-3}$ in the centre\footnote{To avoid unphysically high densities in the interior of the core, we cut off the $r^{-1.5}$-profile at a radius of 0.0125 pc.}. The core is threaded by a magnetic field in the z-direction declining radially outwards with $R^{-0.75}$. It has a strength of 1.3 mG in the centre corresponding to a mass-to-flux ratio of $\mu = 2.6$. The core is rotating rigidly around the z-axis with a rotational energy normalised to the gravitational energy of $\beta_{\rmn{rot}} = 0.04$. Additionally, we add a supersonic turbulence spectrum with a power-law exponent of p = 5/3. The turbulent energy is equal to the rotational energy, i.e. $\beta_{\rmn{turb}} = 0.04$, corresponding to a turbulent rms-Mach number of $\sim 2.5$.

The applied cooling routine~\citep{Banerjee06} takes into account dust cooling, molecular line cooling and the effects of optically thick gas. We introduce sink particles above a density threshold of $\rho_{\rmn{crit}} = 1.14 \cdot 10^{-10}$ g cm$^{-3}$~\citep[see][for details]{Federrath10}. A maximum grid resolution of 1.2 AU is used. The refinement criterion used guarantees that the Jeans length is resolved everywhere with at least 8 grid cells although a even higher resolution has been suggested~\citep{Federrath11}.

To check whether our results are affected by the random turbulent realisation, we performed two more simulations with identical initial conditions but different turbulent seeds. Furthermore, we performed three more simulations with 1) a magnetic field with half the fiducial strength 2) a power-law exponent of the turbulence spectrum of p = 2  and 3) a polytropic cooling to explore the dependency of our results on the initial conditions and numerical methods. All simulations are listed in Table~\ref{tab:models}.
\begin{table}
 \caption{Initial conditions of the performed simulations including run 2.6-4 without turbulence presented in~\citet{Seifried11a}.}
 \label{tab:models}
 \begin{tabular}{@{}lcccccc}
  \hline
  Run & $\mu$  & $\beta_{\rmn{rot}}$ & $\beta_{\rmn{turb}}$ & turbulent seed & p \\
  \hline
  2.6-4-A & 2.6 & 0.04 & 0.04 & A & 5/3 \\
  2.6-4-B & 2.6 & 0.04 & 0.04 & B & 5/3 \\
  2.6-4-C & 2.6 & 0.04 & 0.04 & C & 5/3 \\
  2.6-4-poly & 2.6 & 0.04 & 0.04 & A & 5/3 \\
  2.6-4-A-b & 2.6 & 0.04 & 0.04 & A & 2 \\
  5.2-4-A & 5.2 & 0.04 & 0.04 & A & 5/3 \\
  2.6-4   & 2.6 & 0.04 & 0    & -- & -- \\
  \hline
 \end{tabular}
\end{table}

\section{Results} \label{sec:results}

In total we have performed 6 simulations with varying initial conditions. In the following we present the results of our fiducial run 2.6-4-A in detail. However, we emphasise that the results of the remaining runs are qualitatively very similar. Hence, we are confident that the main findings do not depend on the randomly chosen, initial turbulence field.

\begin{figure}
 \centering
 \includegraphics[width=82mm]{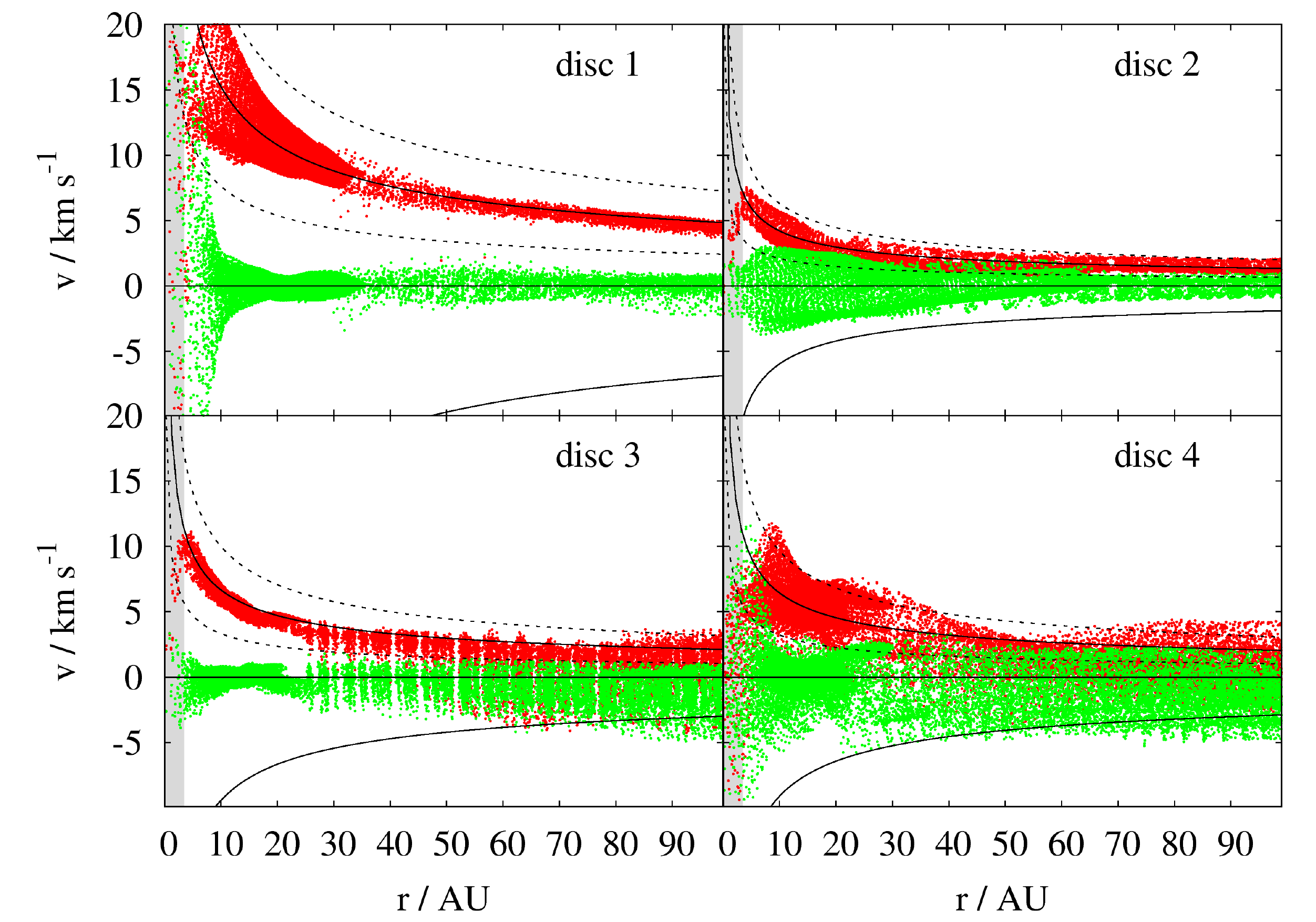}
 \caption{Radial  dependence of the rotation (red) and radial velocity (green) for the four discs formed first in run 2.6-4-A. The black solid line shows the Keplerian velocity $v_{\rmn{kep}}$, the dotted lines 50\% and 150\% of $v_{\rmn{kep}}$. The solid line in the negative velocity range shows the free-fall velocity $v_{\rmn{ff}} = \sqrt{2} v_{\rmn{kep}}$. The regions below 4 AU are affected by resolution effects, therefore they are shaded grey to guide the reader's eye.}
 \label{fig:vel}
\end{figure}

As we are interested in the properties and the evolution of protostellar discs, we restrict our consideration to the time after the first sink particle has formed (t$_0$). Depending on the simulation, this happens after roughly 15 - 20 kyr. At this point large filaments have developed in which the discs form~\citep[for fragmentation and disc formation in massive cores without magnetic fields see][]{Banerjee06b,Girichidis11}.
From this point on we run the simulations for further 10 - 15 kyr. In order to determine global disc properties like mass, centre-of-mass (CoM) and the angular momentum vector we only consider gas with densities larger than $5 \cdot 10^{-13}$ g cm$^{-3}$ around a sink particle. From visual inspection of the density isocontours and variation of this threshold we found this value to be reasonable. Furthermore, it roughly corresponds to the threshold where the gas gets optically thick.  With respect to the CoM and the orientation of the disc we now can calculate the rotation velocity $v_\phi$ and the radial infall velocity $v_{\rmn{rad}}$ for each grid cell less than 20 AU above/below the plane defined by the disc. In order to get an impression of the scatter of these quantities, we do not azimuthally average the values of $v_\phi$ and $v_{\rmn{rad}}$. In Fig.~\ref{fig:vel} we show the radial dependence of $v_\phi$ and $v_{\rmn{rad}}$ 15 kyr after the formation of the first sink particle for the four discs formed first in run 2.6-4-A. Some of the discs have already fragmented and contain more than one sink. The remaining sinks have either no associated discs due to an ejection event caused by many-body interactions or have been created only shortly before the end of run 2.6-4-A so that their discs are not yet well developed, although the velocity profiles reveal Keplerian disc features as well. We note that, in general, the angular momentum vectors of the discs are well off the z-axis. This demonstrates that the discs are created by the local angular momentum associated with turbulent motions and not by the overall rotation of the cloud core. The orientation of the discs does not vary significantly over time as the large-scale structure in which they reside does not change much during the time considered ($\sim$ 10 kyr).
To get an impression of whether the discs are rotationally supported or not, we also plot the Keplerian velocity $v_{\rmn{kep}} = \left( \frac{G M_{\rmn{star}}}{r} \right)^{1/2}$ in Fig.~\ref{fig:vel}, where $G$ is the gravitational constant and $M_{\rmn{star}}$ the mass of all sink particles in the disc. As the disc mass ($\sim$ 0.1 M$_{\sun}$) is significantly smaller than the mass of the sinks we easily can neglect it. As can be seen, $v_\phi$ is close to $v_{\rmn{kep}}$ out to a radius of a few 10 AU (discs 3 + 4) up to 100 AU (discs 1 + 2) with a scatter of about 50\% in each direction as indicated by the dotted lines. This is a remarkable result since for previous simulations of low- and high-mass cores with mass-to-flux ratios $\mu \la 10$ only sub-Keplerian discs were found~\citep[e.g.][]{Allen03,Price07,Mellon08,Hennebelle08,Duffin09,Seifried11a}.

\begin{figure}
 \centering
 \includegraphics[width=37mm]{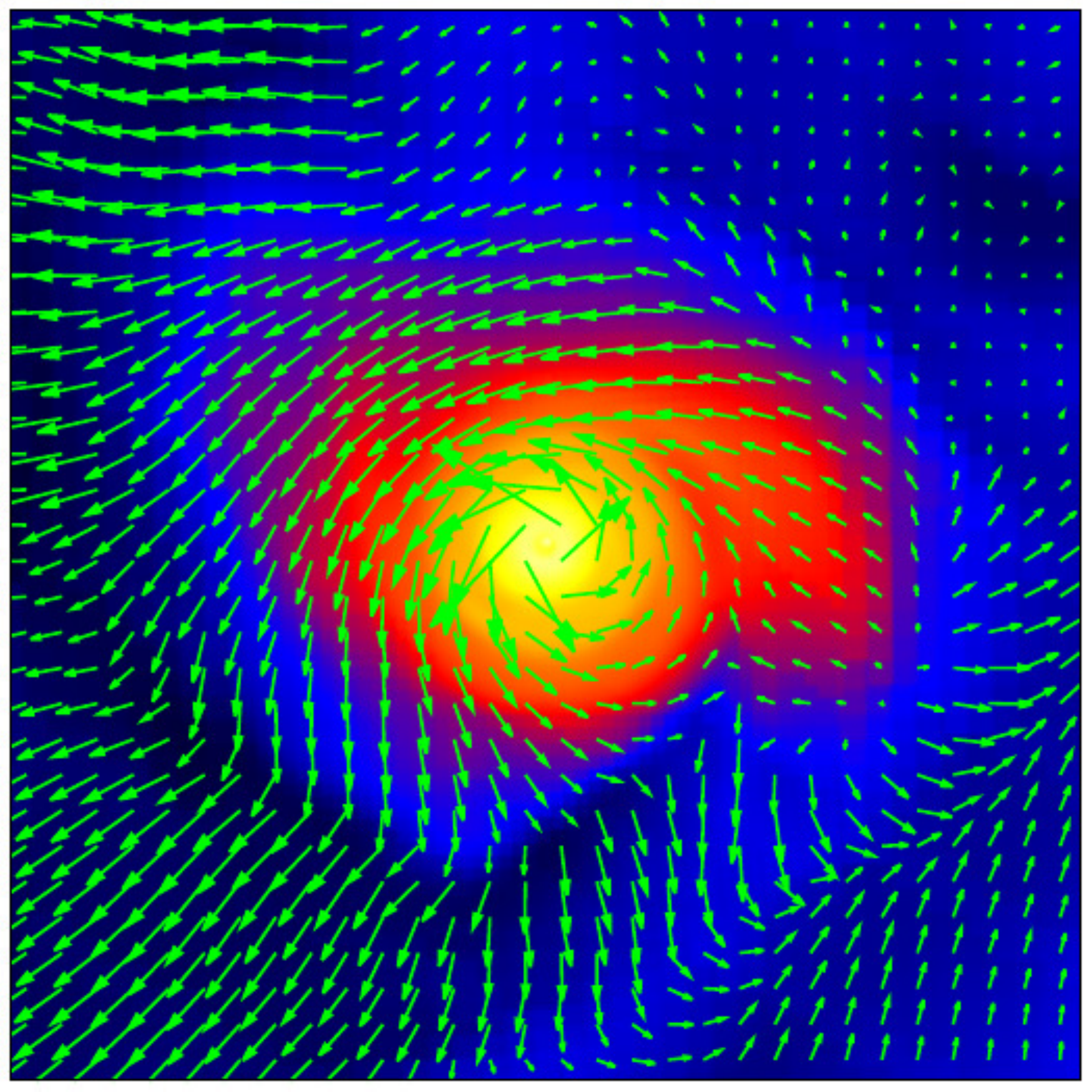}
 \includegraphics[width=37mm]{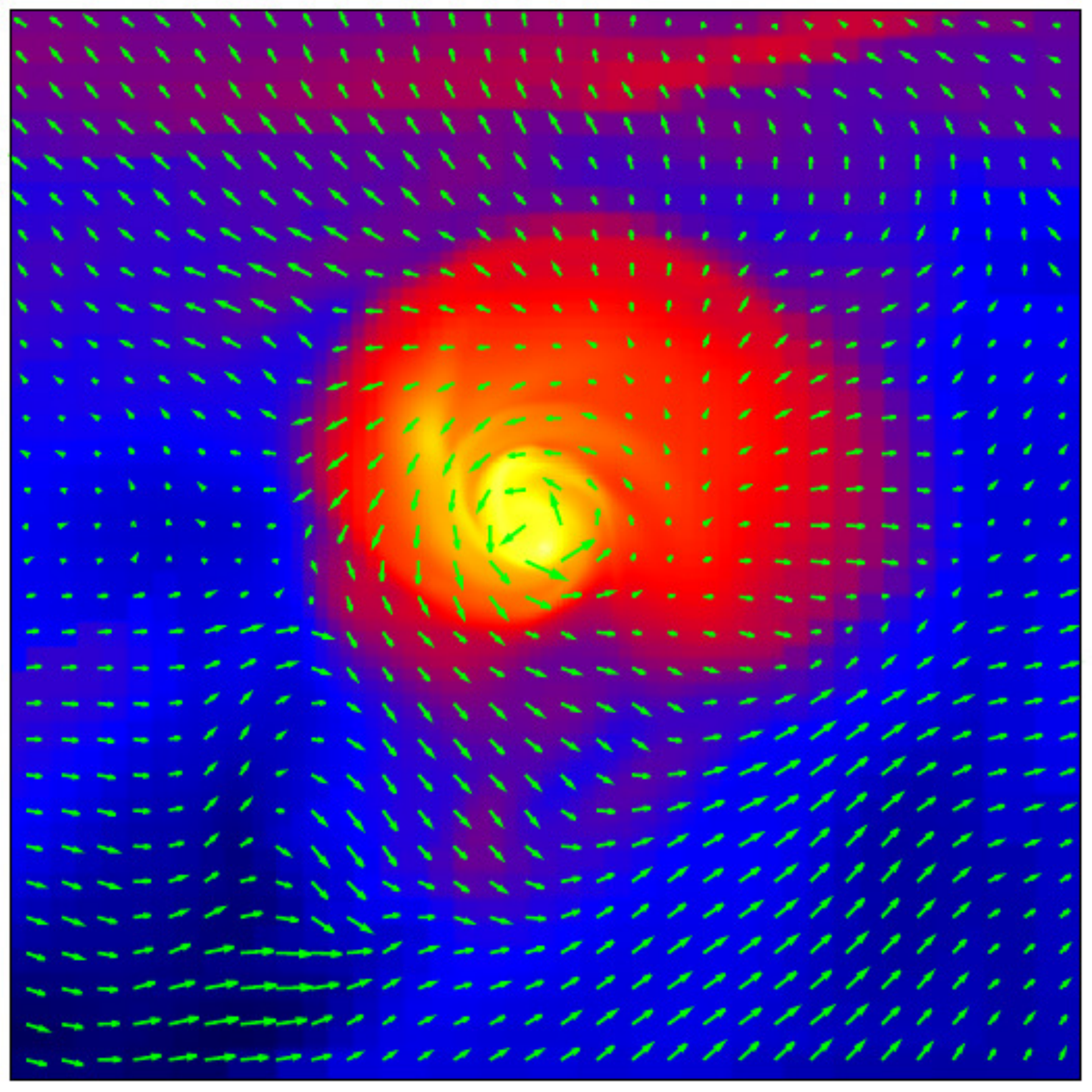} \\
 \includegraphics[width=37mm]{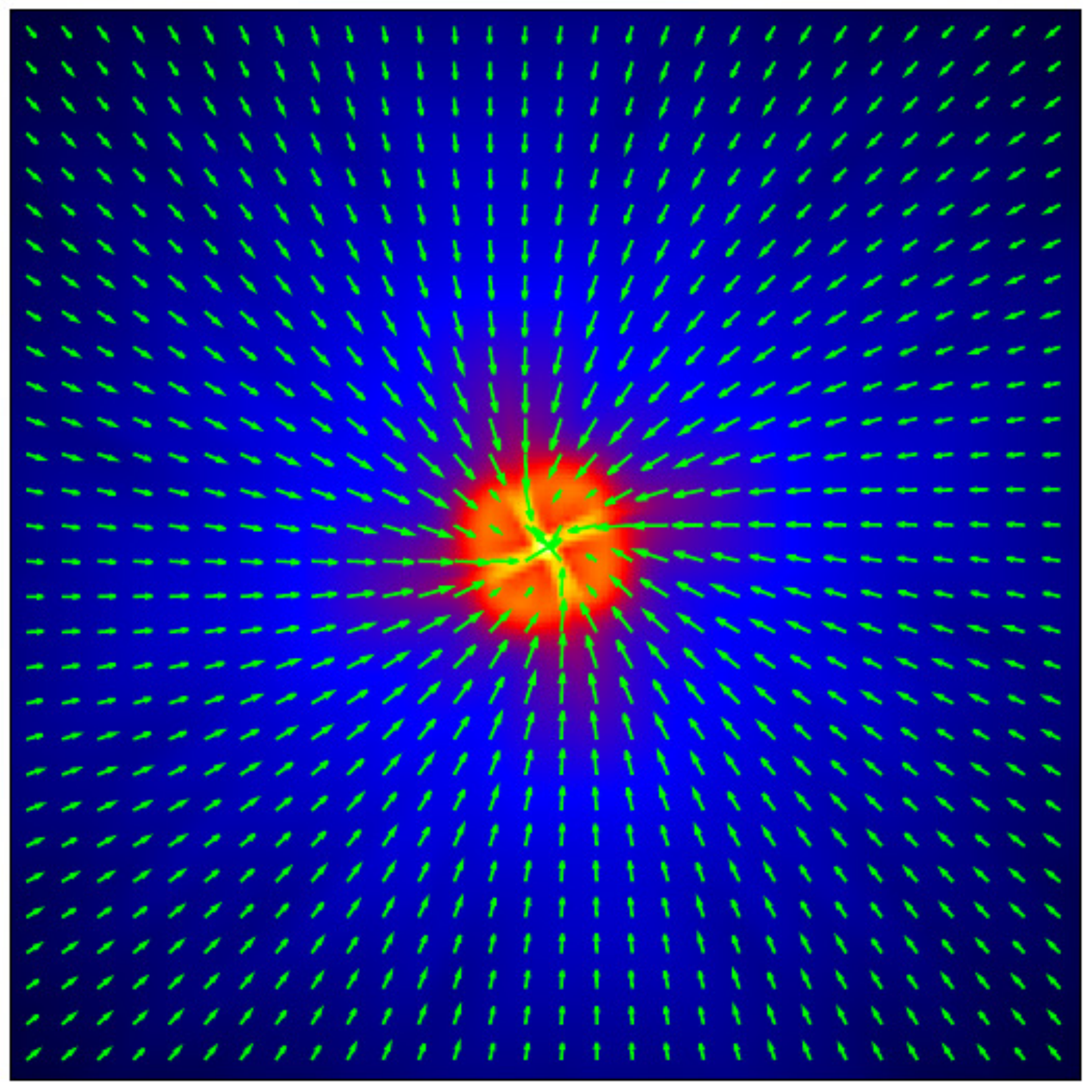}
 \caption{Column density in logarithmic scaling for the top-on view of disc 1 (top left) and disc 2 (top right) of run 2.6-4-A and of the disc in run 2.6-4 without turbulence (bottom). The figures are 800 AU in size.}
 \label{fig:disc}
\end{figure}

We emphasise that for the other runs we find qualitatively similar results, i.e. discs with sizes of up to $\sim$ 100 AU and masses of the order of 0.1 M$_{\sun}$. The number of discs per run varies between 2 and 5. We briefly note that the discs presented here drive molecular outflows. Furthermore, in all discs $v_\phi$ scatters around $v_{\rmn{kep}}$, indicating that this is neither a consequence of the specific turbulence seed (runs 2.6-4-B and 2.6-4-C) nor of the adopted cooling function (run 2.6-4-poly) nor of the power-spectrum exponent (run 2.6-4-b). We find that $v_{\rmn{rad}}$ scatters around 0 and is almost always smaller than $v_\phi$ and significantly smaller than the free-fall velocity $v_{\rmn{ff}} = \sqrt{2} v_{\rmn{kep}}$. This is in strong contrast to the disc in run 2.6-4 which has the same initial setup as the runs presented here except the initial turbulence field~\citep{Seifried11a}. The disc was found to be strongly sub-Keplerian with $v_{\rmn{rad}}$ close to $v_{\rmn{ff}}$. The difference becomes particularly clear when comparing the top-on view of disc 1 and 2 in run 2.6-4-A with that of the disc in run 2.6-4 (Fig.~\ref{fig:disc}).

Why, even in the case of such strongly magnetised cores, are Keplerian discs formed? The suppression of Keplerian disc formation in previous studies without turbulence is due to the very efficient magnetic braking~\citep{Mouschovias80} which removes angular momentum from the midplane at a very high rate. Hence, in our runs the magnetic braking efficiency has to be reduced significantly. Two possible reasons for this are the loss of magnetic flux in the vicinity of the discs or, as proposed recently, a misalignment of the magnetic field and the angular momentum vector~\citep{Hennebelle09,Ciardi10}.

\begin{figure}
 \centering
 \includegraphics[width=84mm]{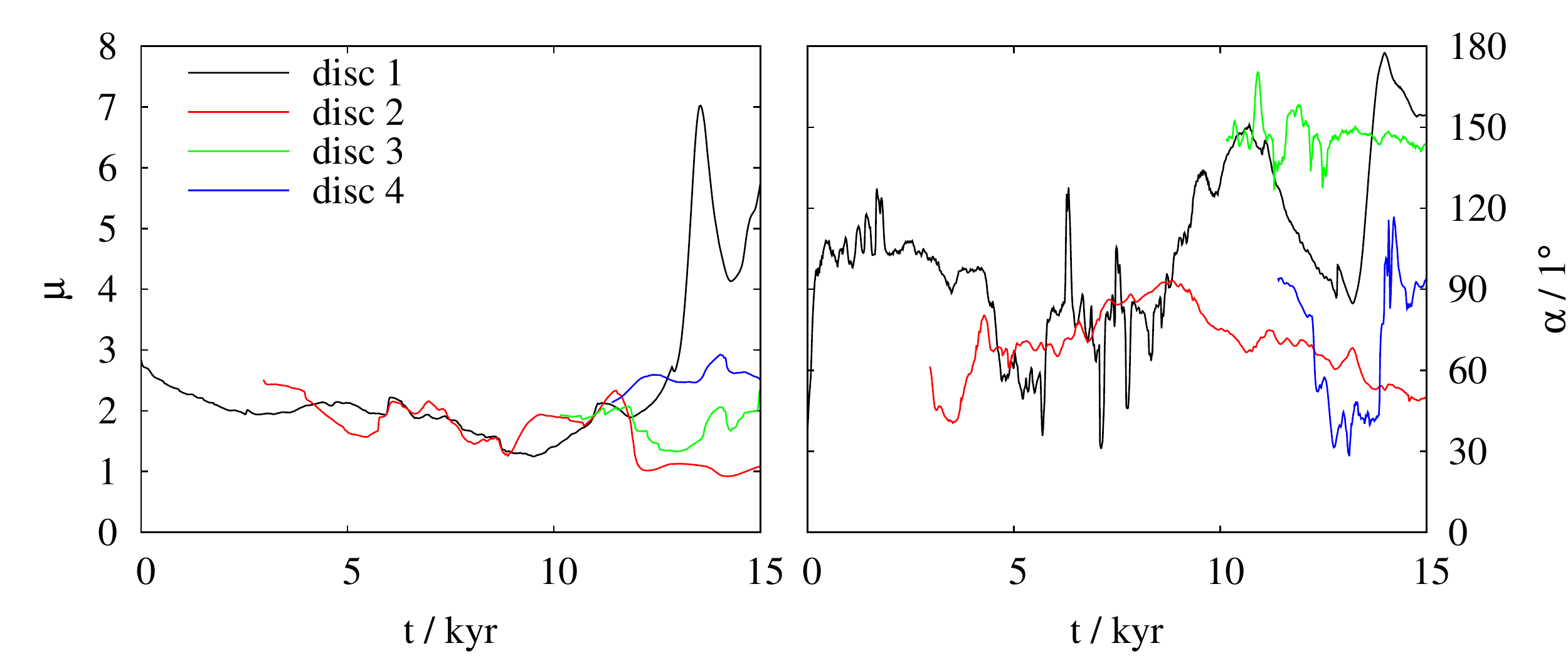}
 \caption{Mass-to-flux ratio $\mu$ (left) and inclination of the mean magnetic field to the angular momentum vector of the disc (right) in spheres with a radius of 500 AU around the CoM of the discs found in run 2.6-4-A.}
 \label{fig:sphere}
\end{figure}

We first consider the possibility of magnetic flux loss in the vicinity of the discs which might be attributed to turbulent reconnection~\citep{Lazarian99}. For this purpose we calculate the volume-weighted, mean magnetic field $\left \langle \mathbfit{B} \right \rangle$ in a sphere with a radius of $r = 500$ AU around the CoM of each disc. In combination with the sphere mass $M$ we obtain the mass-to-flux ratio
\begin{equation}
 \mu = \frac{M}{\pi r^2 |\left \langle \mathbfit{B} \right \rangle|}/\frac{0.13}{\sqrt{G}} \, .
 \label{eq:mu}
\end{equation}
We plot the time variation of $\mu$ in the left panel of Fig.~\ref{fig:sphere} for the same four discs as in Fig.~\ref{fig:vel}.
As can be seen, $\mu$ varies around a mean of 2 - 3. Hence, the values of $\mu$ roughly agree with the overall value of 2.6 and are comparable to the value of $\sim$ 2 found in run 2.6-4. Moreover, $\mu$ is well in the range where simulations without turbulent motions have found sub-Keplerian discs only. We therefore conclude that turbulent reconnection is not responsible for the build-up of Keplerian discs in our runs.

Another way of reducing the magnetic braking efficiency was investigated by~\citet{Hennebelle09} and~\citet{Ciardi10}. These authors found that even for a small misalignment of the overall magnetic field and the rotation axis Keplerian discs can form. As we consider a turbulent flow, it is very likely that the magnetic field and the rotation axis are misaligned. In the right panel of Fig.~\ref{fig:sphere} we plot the angle $\alpha$ between the disc angular momentum vector and $\left \langle \mathbfit{B} \right \rangle$ in the spheres around the discs of run 2.6-4-A. The angle $\alpha$ is significantly larger than 0$^{\circ}$ which supports the picture of a reduced magnetic braking efficiency due to a misalignment of the magnetic field and the rotation axis.

However, there is a third way to reduce the magnetic braking efficiency while simultaneously keeping the inwards angular momentum transport on a high level. Considering the top panel of Fig.~\ref{fig:disc} it can be seen that in the surroundings of each disc there is a turbulent velocity field with no signs of a coherent rotation structure. Therefore no toroidal magnetic field component (w.r.t. the coordinate system of the disc) can be built up. But as the angular momentum is mainly extracted by toroidal Alfv\`enic waves, it is not surprising that the magnetic braking efficiency is strongly reduced in the environment of the disc despite a low mass-to-flux ratio (compare left panel of Fig.~\ref{fig:sphere}). Despite the lack of a coherent rotation structure, locally the inwards angular momentum transport can remain high due to \textit{local} shear flows driving large angular momentum fluxes. We also note that the non-coherent flow cannot be efficiently slowed down by the magnetic field as it does in case of large-scale coherent motions. This can be seen in our previous simulations~\citep{Seifried11a} without initial turbulence. Here the angular momentum is removed almost completely \textit{before} the gas hits the disc (see also bottom panel of Fig.~\ref{fig:disc}). Hence, it is the shear flow generated by turbulent motions that leads to the build-up of Keplerian discs.

To quantify this, we calculate the torques of the gas $\tau_{\rmn{gas}}$ and the magnetic field $\tau_{\rmn{mag}}$ in cylinders of variable radii and a total height of 40 AU\footnote{This height is found to be reasonable by visually inspecting the discs in edge-on view. Furthermore, a variation of this value does not qualitatively change the overall picture.} around the CoM of each disc. The symmetry axis of the cylinders is determined by the angular momentum vector of the corresponding disc.
\begin{figure}
 \includegraphics[width=62mm]{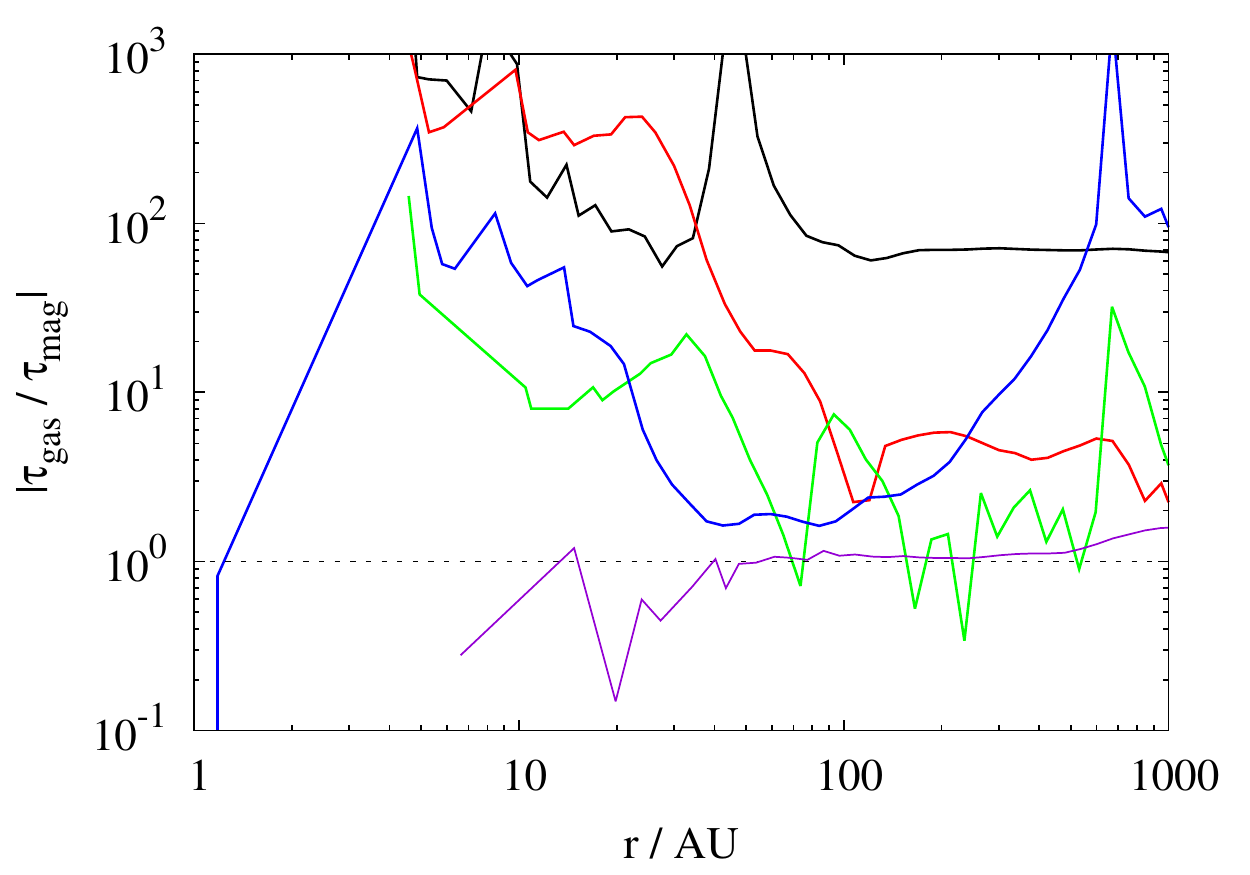}
 \caption{Ratio of $\tau_{\rmn{gas}}$ to $\tau_{\rmn{mag}}$ for the discs found in run 2.6-4-A (same colours as in Fig.~\ref{fig:sphere}). and for the disc in run 2.6-4 without turbulence (violet line).}
 \label{fig:torques}
\end{figure}
In Fig.~\ref{fig:torques} we plot the ratio of $\tau_{\rmn{gas}}$ to $\tau_{\rmn{mag}}$ for the four discs found in run 2.6-4-A. On average, $\tau_{\rmn{gas}}$ exceeds $\tau_{\rmn{mag}}$ by at least a factor of a few, i.e. angular momentum is transported inwards at a higher rate than it is extracted by the magnetic field. This is also observed for the discs in the other runs not shown here.
We note that the gravitational torques not shown here are generally even smaller than $\tau_{\rmn{mag}}$. The only exception occurs for disc 2 in run 2.6-4-A where a strong spiral arm has formed exerting a strong gravitational torque (see right panel of Fig.~\ref{fig:disc}). We also note that the strong fluctuations of $\tau_{\rmn{gas}}/\tau_{\rmn{mag}}$ for disc 3 (green line) are due to strong perturbations in its vicinity. Disc 4, for example, perpendicular intercepts the plane defined by disc 3 causing the drop at $r \simeq$ 90 AU.

In general, however, there is a net transport of angular momentum inwards resulting in the observed build-up of Keplerian discs. In contrast, for the non-turbulent run $\tau_{\rmn{gas}}$ is almost perfectly balanced by $\tau_{\rmn{mag}}$. Furthermore, comparing the absolute values of $\tau$ between the turbulent and non-turbulent case shows that -- although this comparison is somewhat crude -- in general in the turbulent case $\tau_{\rmn{mag}}$ is reduced significantly whereas $\tau_{\rmn{gas}}$ remains comparable.

This fits in the picture described above where the magnetic braking efficiency is reduced due to the absence of a coherent rotation structure and the angular momentum transport by the gas remains high due to local shears flows. Hence, this confirms our assumption that the turbulent disc environment is responsible for the build-up of Keplerian discs.

We also made a rough estimate for the magnetic braking time $T_{\rmn{mag}} = \rho_{\rmn{disk}}/\rho_{\rmn{env}} \cdot Z/v_{\rmn{Alf}}$ in the discs themselves, calculating the mean density $\rho_{\rmn{disk}}$ and the scale height $Z$ of the discs as a function of radius. Assuming a density contrast $\rho_{\rmn{disk}}/\rho_{\rmn{env}} = 100$ shows that for the discs $T_{\rmn{mag}}$ is on average a factor of 5 - 10 above the characteristic timescale of inwards angular momentum transport $L_{\rmn{disk}}/\tau_{\rmn{gas}}$. However, we again emphasise that the magnetic braking efficiency has to be reduced in the region \textit{external} to the discs in order to allow for the build-up of Keplerian discs.

\section{Discussion}

\begin{figure}
 \centering
 \includegraphics[width=62mm]{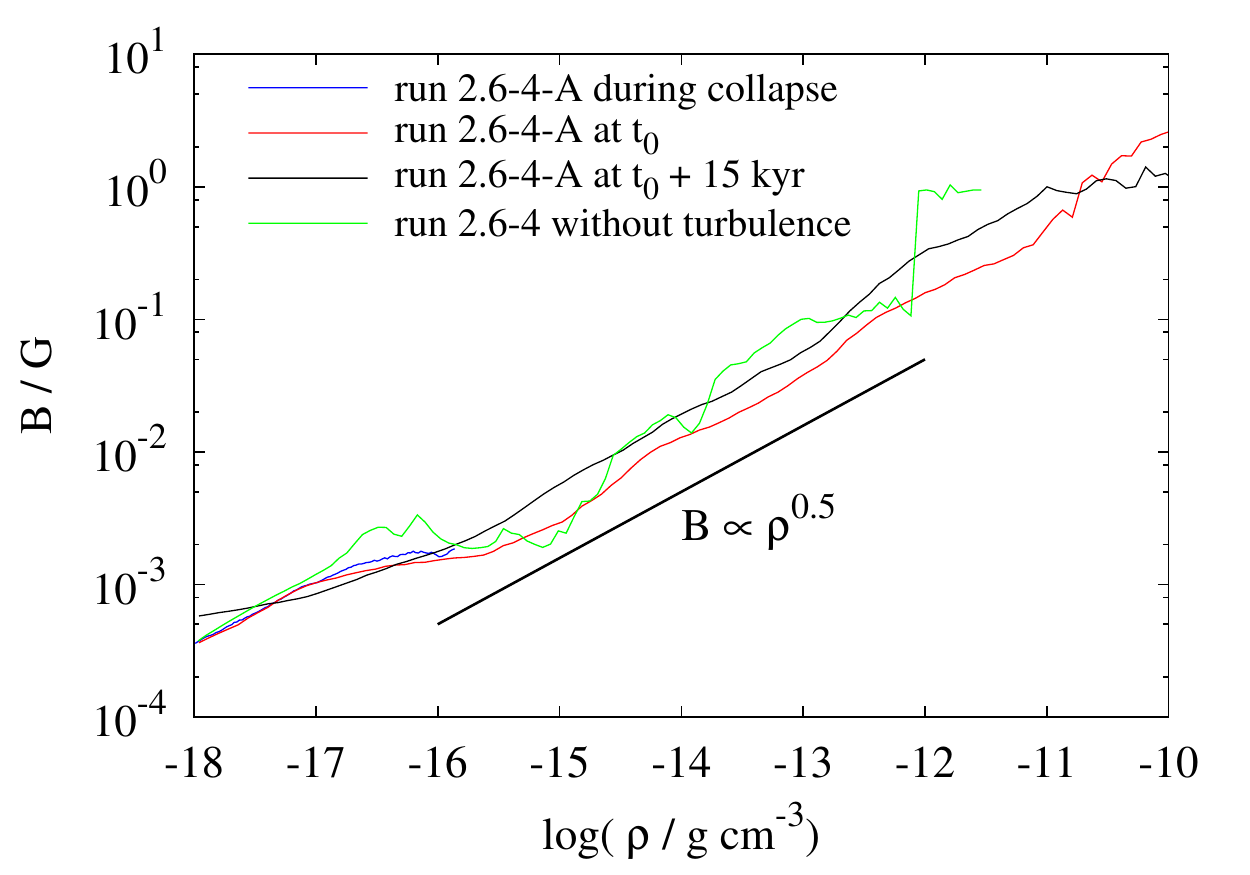}
 \caption{Scaling of the magnetic field for different times in run 2.6-4-A as well as in run 2.6-4.}
 \label{fig:scaling}
\end{figure}

Recently \citet{Santos12} have reported the formation of Keplerian discs in turbulent, strongly magnetised low-mass cores. These authors, however, attribute this to the effect of turbulent reconnection~\citep{Lazarian99} lowering the magnetic flux. As shown in the left panel of Fig.~\ref{fig:sphere}, we find the value of $\mu$ on small scales to be comparable with the overall value, indicating that there is no magnetic flux loss. To further support this, we plot the scaling of the magnetic field with the density in Fig.~\ref{fig:scaling}.
The observed scaling $B \propto \rho^{0.5}$ is very similar to the non-turbulent case~\citep[run 2.6-4 in][]{Seifried11a} and does not change significantly over time. Hence, we argue that in our case no significant turbulent reconnection occurs and that magnetic flux loss is not responsible for the formation of Keplerian discs as proposed by \citet{Santos12}. Further comparison to their work is not possible as the authors do not consider the scaling of the magnetic field.

We note that in some cases $\mu$ (Eq.~\ref{eq:mu}) slightly increases with decreasing radius and eventually reaches values above 10 at radii $\la$ 100 AU. Hence, it could be argued that turbulent reconnection is happening on these scales. Also from Fig.~\ref{fig:scaling} some flux loss is apparent at densities above $\sim$ 10$^{-11}$ g cm$^{-3}$. Typical scales associated with this density are of the order of 30 - 40 AU. However, on these scales ($\la$ 100 AU) the velocity structure seems to be relatively well ordered (Fig.~\ref{fig:disc}) which makes turbulent reconnection unlikely to happen. Hence, we rather attribute the flux loss to numerical diffusion. However, we again point out that, for enough angular momentum being available to form a Keplerian disc on scales of $\sim$ 100 AU, already \textit{on larger scales} ($\ga$ 500 AU) the magnetic braking has to be reduced. On such large scales, however, no significant magnetic flux loss is observed (see left panel of Fig.~\ref{fig:sphere}). Hence, whatever accounts for the flux loss on scales $\la$ 100 AU, it does not affect the mechanism postulated in this work.

The large misalignment of the magnetic field of the spheres and the angular momentum of the discs of up to 90$^\circ$ (see Fig.~\ref{fig:sphere}) also might weaken the effect of magnetic braking as suggested by \citet{Hennebelle09} and \citet{Ciardi10}. In their work, however, uniformly rotating spheres were considered which is clearly not the case here (see Fig.~\ref{fig:disc}). Therefore, it is hard to tell to what extent in our case the misalignment affects the magnetic braking efficiency. Our disc analysis shows that the misalignment is not the main reason for the decreased magnetic braking efficiency. Here we see that the angular momentum is ``generated'' locally by turbulent shear flows and that the classical picture of magnetic braking for coherently rotating structure breaks down leading to the build-up of 
Keplerian discs.

Recently, \citet{Krasnopolsky11} have proposed that including the Hall effect can result in the formation of large-scale Keplerian discs. They claim, however, that a Hall coefficient about one order of magnitude larger than expected under realistic conditions would be required. Also Ohmic dissipation fails to produce Keplerian discs larger than roughly 10 solar radii in the early evolutionary stages~\citep{Dapp11a,Dapp11b}, unless a strongly enhanced resistivity is used~\citep{Krasnopolsky10}. Hence, it seems that all three non-ideal MHD effects~\citep[see][for the case of ambipolar diffusion]{Mellon09} have a hard time in accounting for the formation of Keplerian discs. However, as we have shown, already for the ideal MHD limit Keplerian discs can form in strongly magnetised cores when turbulent motions are included. Therefore it seems that non-ideal MHD effects or turbulent reconnection are not necessarily required to avoid catastrophic magnetic braking.

Earlier observations of individual Class 0 sources found cases where no well defined, Keplerian disc were detected~\citep[e.g.][]{Belloche02}. These observations suggest that discs might form at later stages which would alleviate the magnetic braking problem. However, comparing more recent observations of samples of Class 0 sources with detailed radiative transfer models suggests that the majority of them harbours well defined protostellar discs forming in the earliest stages~\citep{Jorgensen09,Enoch11}.

We note that recently~\citet{Hennebelle11} and \citet{Commercon11} also performed simulations of high-mass turbulent cloud cores finding large-scale outflows which require the existence of protostellar discs supporting our arguments here.

\section{Conclusion} \label{sec:conclusion}

We have performed collapse simulations of strongly magnetised ($\mu = 2.6, 5.2$), 100 M$_{\sun}$ cloud cores. A turbulent velocity field was superposed on the uniform core rotation. We find that after an initial collapse phase of $\sim$ 15 kyr discs with typical masses of $\sim$ 0.1 M$_{\sun}$ form. The discs are up to 100 AU in size and have Keplerian rotation velocities, a result in strong contrast to previous simulations of strongly magnetised cores containing no initial turbulence. We showed that our findings neither depend on the randomly chosen turbulence field nor on the adopted cooling routine.

We suggest that the main reason for Keplerian disc formation is the turbulent surroundings of the disc. As there is no coherent rotation structure on scales of several 100 AU, the generation of a toroidal magnetic field is suppressed therefore lowering the magnetic braking efficiency already \textit{before} the gas hits the disc. At the same time the inwards angular momentum transport by the gas remains high due to local shear flows which results in a net inwards angular momentum transport.

Our work strongly suggests that the ``magnetic braking catastrophe'' as reported in numerous papers is more or less a consequence of the highly idealised initial conditions neglecting turbulent motions. A turbulent velocity structure in the surroundings of the disc, as obtained with more realistic initial conditions, is enough to allow for the formation of Keplerian discs. Other effects like misaligned magnetic fields and angular momentum vectors, turbulent reconnection or non-ideal MHD effects seem not to be necessary. Turbulence alone provides a natural and at the same time very simple mechanism to solve the ``magnetic braking catastrophe'' problem.

\section*{Acknowledgements}

The authors like to thank the anonymous referee for his comments which helped to significantly improve the paper. D.S. and R.B. acknowledge funding of Emmy-Noether grant 3706/1-1 by the DFG. R.E.P is supported by a Discovery grant from NSERC of Canada. R.S.K. acknowledges subsidies from the {\em Baden-W\"urttemberg-Stiftung} (grant P-LS-SPII/18) and from the German {\em Bundesministerium f\"ur Bildung and Forschung} via the ASTRONET project STAR FORMAT (grant 05A09VHA). R.S.K. furthermore gives thanks for subsidies from the Deutsche Forschungsgemeinschaft (DFG) under grants no.\ KL 1358/11 and KL 1358/14 as well as via the Sonderforschungsbereich SFB 881 {\em The Milky Way System}.
The simulations were performed on HLRB2 at the Leibniz Supercomputing Centre in Garching and on JUROPA at the Supercomputing Centre in J\"ulich. 
The FLASH code was developed partly by the DOE-supported Alliances Center for Astrophysical Thermonuclear Flashes (ASC) at the University of Chicago.

\label{lastpage}

\end{document}